\documentclass[titlepage, a4paper, 10pt, twocolumn]{article}
\usepackage{graphicx}
\usepackage{setspace}
\usepackage[margin=1in]{geometry}
\usepackage{multicol}
\usepackage[english]{babel}
\usepackage{fancyhdr,lastpage}
\usepackage{amsfonts}
\usepackage{amsthm}
\usepackage{amsmath}
\usepackage{color}
\usepackage{float}
\usepackage{url}
\usepackage[font=small]{caption}

\begin{document}

\bibliographystyle{vancouver}

\singlespacing

\onecolumn

\section*{MAGNA: Maximizing Accuracy in Global Network Alignment}
\textbf{Vikram Saraph and Tijana Milenkovi\'{c}$^*$}\\
\textbf{Department of Computer Science and Engineering}\\
\textbf{University of Notre Dame, Notre Dame, IN 46556,  USA}\\
$^*$Corresponding Author (E-mail: tmilenko@nd.edu)

\section*{Summary}

Biological network alignment aims to identify similar regions between
networks of different species.  Existing methods compute node
``similarities'' to rapidly identify from possible alignments the
``high-scoring'' alignments with respect to the overall node
similarity.  However, the accuracy of the alignments is then evaluated
with some other measure that is different than the node similarity
used to construct the alignments.  Typically, one measures the amount
of conserved edges.  Thus, the existing methods align similar
\emph{nodes} between networks \emph{hoping} to conserve many
\emph{edges} (\emph{after} the alignment is constructed!).

Instead, we introduce MAGNA to directly ``optimize'' edge conservation
\emph{while} the alignment is constructed.  MAGNA uses a genetic
algorithm and our novel function for ``crossover'' of two ``parent''
alignments into a superior ``child'' alignment to simulate a
``population'' of alignments that ``evolves'' over time; the
``fittest'' alignments survive and proceed to the next ``generation'',
until the alignment accuracy cannot be optimized further.  While we
optimize our new and superior measure of the amount of conserved
edges, MAGNA can optimize \emph{any} alignment accuracy measure.  In
systematic evaluations against existing state-of-the-art methods
(IsoRank, MI-GRAAL, and GHOST), MAGNA improves alignment
accuracy of \emph{all} methods.

\bigskip
\medskip

\twocolumn

\section{Introduction}

\subsection{Motivation and background}\label{sect:motivation}

Genomic sequence alignment has led to breakthroughs in our
understanding of how cells work.  It identifies regions of similarity
between sequences of individual genes that are a likely consequence of
evolutionary relationships between the sequences.  However, genes,
i.e., their protein products, do not act alone but instead interact
with each other to carry out cellular processes.  And this is exactly
what \emph{protein-protein interaction (PPI) networks} model. (While
we focus on PPI networks, our ideas are applicable to \emph{any}
network type.) Then, network alignment (NA) can be used to find
regions of similarities between PPI networks of different species that
are a likely consequence of evolutionary relationships between the
networks.  Unlike sequence alignment that ignores genes'
interconnectivities, NA allows for studying complex cellular events
that are a consequence of the \emph{collective} behavior of the genes'
protein products.  As such, NA is promising to \emph{further} our
biological understanding \cite{Sharan2006}.

As recent biotechnological advances continue to yield large amounts of
PPI data \cite{BIOGRID}, alignment of PPI networks of different
species continues to gain importance \cite{Sharan2006}. This is
because NA could guide the transfer of biological knowledge across
species between conserved (aligned) network regions \cite{Sharan2006}.
This is important, since many proteins remain functionally
uncharacterized even for well studied species \cite{Sharan2007}.
Traditionally, the across-species transfer of biological knowledge has
relied on sequence alignment.  However, since PPI networks and
sequences can capture complementary functional slices of the cell,
implying that PPI networks can uncover function that cannot be
uncovered from sequences by current methods, restricting alignment to
sequences may limit the knowledge transfer \cite{Memisevic10b}.

Unfortunately, the mathematics of complexity theory dictates that
exact network (or graph) comparison is computationally intractable.
The underlying problem is that of subgraph isomorphism, which asks
whether one graph (the source) appears as an exact subgraph of another
graph (the target). Answering this is NP-complete \cite{Cook1971}.
Furthermore, simply answering the subgraph isomorphism problem is not
enough when comparing PPI networks, since one PPI network is rarely an
exact subnetwork of another due to biological variation \cite{GRAAL}.
It is much more desirable to answer how similar two networks are and
in what regions they share similarity.  NA can be used for this
purpose.

NA is a less restrictive problem than that of subgraph isomorphism, as
it seeks to ``fit'' the source into the target in the ``best possible
way'' even if the source is not an exact subgraph of the target.  An
alignment is a mapping between nodes of the source and nodes of the
target that is expected to conserve as much structure (or topology) as
possible between the two networks.  (Note that methods exist that can
align more than two networks \cite{IsoRankN,Flannick2008}, but we
focus on \emph{pairwise} NA.)  Since NA is computationally hard,
heuristic methods must be sought.

NA can be \emph{local} (LNA) or \emph{global} (GNA).  Initial
solutions for NA have aimed to match local network regions
\cite{PathBlast,Sharan2005,Flannick2006,Mawish,Berg04,Liang2006a,Berg2006}.
That is, in LNA, subnetworks, rather than the entire networks, are
aligned. However, aligned regions can overlap, leading to
``ambiguous'' many-to-many node mappings.  Thus, GNA solutions have
been proposed
\cite{Singh2007,Flannick2008,Singh2008,GraphM,IsoRankN,GRAAL,HGRAAL,MIGRAAL,GHOST,Narayanan2011,NATALIE,NETAL}.
In contrast to LNA, GNA compares entire networks, typically by
aligning every node in the source to exactly one unique node in the
target.  We focus on \emph{GNA}, but our ideas are also applicable to
LNA.

Traditionally, GNA has relied on biological information
\emph{external} to network topology, e.g., sequence similarity
\cite{Sharan2006}.  To extract the most from each source of biological
information, it would be good to know how much of new biological
knowledge can be uncovered \emph{solely} from network topology
\emph{before} integrating it with other sources of biological
information \cite{GRAAL,HGRAAL,MIGRAAL,GHOST,NETAL}.  Only after
methods for topological GNA are developed that result in alignments of
good topological \emph{and} biological quality, it would be beneficial
to integrate them with other biological data sources to further
improve the quality. Thus, we focus on \emph{topological} GNA, but
additional biological data can easily be added.

Existing GNA methods, of which the more prominent ones (and which we
consider in our study) are outlined below, typically use a two-step
approach: (1) score the ``similarity'' of pairs of nodes from
different networks, and (2) feed these scores into an alignment
strategy to identify ``high-scoring'' alignments from all possible
alignments.

\emph{IsoRank} \cite{Singh2007} scores nodes from two networks by a
PageRank-based spectral graph theoretic principle: two nodes are a
good match if their neighbors are good matches.
After these topological scores are computed, biological scores can be
added to get final node scores.  An alignment is then constructed by
greedily matching the high-scoring node pairs. IsoRank has evolved
into \emph{IsoRankN} to allow for \emph{multiple} GNA \cite{IsoRankN}
and \emph{many-to-many} node mapping, but this is out of the scope of
our study.

The \emph{GRAAL} family of algorithms
\cite{GRAAL,HGRAAL,CGRAAL,MIGRAAL}, developed in parallel with the
IsoRank family, use graphlet (or small induced subgraph) counts to
compute mathematically rigorous topological node similarity scores
\cite{Milenkovic2008,MMGP_Roy_Soc_09,Solava2012}.  Intuitively, two
nodes are a good match if their \emph{extended} network neighborhoods
are ``topologically similar'' with respect to the graphlet counts.
Also, \emph{MI-GRAAL} \cite{MIGRAAL}, the latest of the family
members, can automatically add other (biological) node similarity
scores into final scores.  It is the alignment strategies of the GRAAL
family members that are different.  MI-GRAAL combines alignment
strategies of the other members, thus outperforming each of them
\cite{MIGRAAL}.

More recent \emph{GHOST} \cite{GHOST} uses ``spectral signatures'' to
score node pairs topologically while also allowing for inclusion of
biological node scores.  Similar to MI-GRAAL, GHOST's alignment
strategy is also seed-and-extend, except that GHOST solves a quadratic
assignment problem, while MI-GRAAL solves a linear assignment problem.

\subsection{Our contribution}

Recall that the existing GNA methods construct alignments by scoring
all node pairs with respect to the nodes' similarities and by rapidly
identifying ``high-scoring'' alignments from all possible alignments.
Here, ``high-scoring'' alignments are typically those that
``maximize'' (greedily or optimally) the node similarity score totaled
over all mapped nodes \cite{Singh2007,GRAAL,HGRAAL,MIGRAAL,GHOST}.
However, the accuracy (or quality) of the alignments is then evaluated
with respect to some other measure of an inexact fit of two networks,
which is different than the node scoring function that is used to
construct the alignments in the first place.  Typically, one measures
the amount of conserved edges (see below) \cite{MIGRAAL,GHOST}. (One
also evaluates the alignments biologically with respect to functional
knowledge.)  Thus, the existing methods align ``similar'' \emph{nodes}
between networks with the goal (or \emph{hope}!) of conserving as many
\emph{edges} as possible under the alignment (\emph{after} the
alignment is constructed!).  Instead, we introduce \emph{MAGNA}, a new
framework for directly ``\textbf{m}aximizing'' (or ``optimizing'')
\textbf{a}ccuracy in \textbf{GNA} with respect to the amount of
conserved edges \emph{while} the alignment is being constructed, which
is our first contribution.

Optimizing the amount of conserved edges would require finding a
global optimum over the search space consisting of all possible node
mappings. Due to the large size of the space, exhaustive search is
computationally intractable. But, approximate techniques exist with
solutions very close to optimal, such as genetic algorithms
\cite{Cross2000}. Hence, we adapt the idea of genetic algorithms to
the problem of GNA to develop MAGNA as a conceptually novel GNA
framework. This is our second contribution, since to our knowledge,
genetic algorithms have not been used for PPI GNA thus far.  MAGNA
simulates a ``population'' of alignments that ``evolves'' over time
(the initial population can consist of random alignments or of
alignments produced by the existing methods).  Then, the ``fittest''
candidates (those that conserve the most edges)
survive and proceed to the next generation. This is repeated until the
algorithm converges, i.e., until the amount of conserved edges cannot
be optimized further.

Much of what defines any genetic algorithm is the crossover function,
which ``combines'' two candidates (i.e., alignments) into a new one.  And
since genetic algorithms have not been used for GNA thus far, we had
to devise a novel (and to our knowledge, \emph{the first ever})
function for crossover of two parent alignments into a child alignment
that reflects (ideally the best of) each parent.  The alignment
crossover function is the third and a major contribution of our study,
because it allows MAGNA not only to combine alignments produced by
\emph{any} existing method to \emph{improve} them but also to produce
\emph{its own new} superior alignments.

It is not obvious how to measure the quality of an alignment
\cite{HGRAAL}, i.e., which measure to optimize as the ``fitness''
function within the genetic algorithm. Clearly, a good alignment
should maximize the amount of conserved edges. Different measures have
been proposed to quantify this, all of which are heuristics and thus
correctly reflect the actual alignment quality in some cases but fail
to do so in other cases.  Thus, as our fourth contribution, we
introduce a new and superior alignment quality measure that takes the
best from each existing measure.  While we optimize with MAGNA this new
measure as well as the existing measures of the amount of conserved
edges, importantly, MAGNA can optimize \emph{any} measure of alignment
quality, topological \emph{or} biological, which is another of our
contributions.

We evaluate MAGNA against IsoRank, MI-GRAAL, and GHOST by aligning
a high-confidence yeast PPI network with its noisy counterparts, where
the true node mapping is known.  This popular
evaluation test \cite{GRAAL,MIGRAAL,GHOST} allows for a systematic
method comparison. MAGNA improves alignment
quality of \emph{all} of the existing methods.

\section{Methods}

\subsection{Alignment crossover function}\label{sect:crossover}

In this section, we provide the mathematical rigor necessary to define
our novel ``crossover function'', which is at the heart of MAGNA.  (The
description of MAGNA is given in Section \ref{sect:MAGNA}.)  The
crossover function should take two ``parent'' alignments and produce a
``child'' alignment that is intended to reflect (ideally the best of)
both parents.

Let $G_1(V_1, E_1)$ and $G_2(V_2, E_2)$ be two networks with $V_i$ and
$E_i$ as the sets of nodes and edges, respectively. Let $m = |V_1|$
and $n = |V_2|$.  Without loss of generality, suppose $|V_1|
\le |V_2|$. An \emph{alignment} of $G_1$ to $G_2$ is a total injective
function $f : V_1 \rightarrow V_2$. That is, every element of $V_1$ is
matched uniquely with an element of $V_2$. If $|V_1| = |V_2|$, when
$f$ is an injective function, then in fact $f$ is a bijection.

Let $V_1 = \{x_1, \ldots, x_m\}$ and $V_2 = \{y_1, \ldots, y_n\}$.
Let $[n] = \{1, \ldots, n\}$ be the set of natural numbers from 1 to
$n$. A \emph{permutation} is a bijection $\sigma : [n] \rightarrow
[n]$. Then, with the assumption that $m = n$, and given this fixed
number labeling of nodes as above, we can represent any alignment $f$
with a corresponding permutation $\sigma$ that maps node labels to
node labels. Even though it is rare that $|V_1| = |V_2|$, we can
easily force this condition to be true by adding dummy, zero-degree
nodes $z_i$ to $V_1$, as follows: $\bar{V_1} = V_1 \cup \{z_{m+1},
\ldots, z_n\}$. Thus, from now on, we will simply assume that $|V_1| =
|V_2|$, without explicitly referring to $\bar{V_1}$.  Therefore, any
alignment can be represented as a permutation, and for the remainder
of this section, we use ``permutation'' and ``alignment''
interchangeably. This representation is critical to our crossover
function.

Let $S_n$ denote the set of all permutations. Notice that $|S_n| =
n!$, which is large. In theory, to find an alignment of maximum
quality (with respect to a given criterion), we could ``simply''
enumerate all permutations and evaluate the quality of each one.
However, this is impractical due to the large size of $S_n$, so we
require a clever search heuristic.  We design such a heuristic as
follows. 
First, we create a graph with $S_n$ as the set of nodes in which two
permutations (alignments) are connected by an edge if the alignments
are ``adjacent'' (see below). Second, since intuitively the alignment
quality is continuous in alignment ``adjacency'' (in the sense that
two ``adjacent'' alignments should be of similar quality, or in other
words, a small perturbation of an alignment should not greatly affect
its quality), we exploit the topology of this graph to define a
function for crossover of two alignments. Namely, we define the child
alignment as the alignment that is ``in the middle'' between two given
parent alignments in this graph.  Formal details are as follows.

Given two permutations $\sigma$ and $\tau$, we define what it means
for $\sigma$ and $\tau$ to be adjacent. A \emph{transposition} of a
permutation is a new permutation that fixes every element of the
original permutation, except two elements, which are swapped.  Then,
two permutations are \emph{adjacent} if they differ by a
transposition; that is, $\sigma$ and $\tau$ are adjacent if there is a
transposition $\rho$ such that $\sigma = \rho \circ \tau$. We create
graph $\Gamma_n$ with the set of nodes $S_n$ and the set of edges
$E_n$, where an edge between $\sigma$ and $\tau$ is in $E_n$ if and
only if $\sigma$ and $\tau$ are adjacent.  Then, we define $\sigma
\otimes \tau$, the \emph{crossover} of \emph{any} two permutations
$\sigma$ and $\tau$ from $S_n$, as a permutation which is the midpoint
on a shortest path from $\sigma$ to $\tau$ in $\Gamma_n$.  This
definition captures what we desire from a crossover function. More
precisely, it can be shown that for randomly selected permutations
$\sigma$ and $\tau$, $|\sigma \cap (\sigma \otimes \tau)| / n
\rightarrow 1/2$ and $|\tau \cap (\sigma \otimes \tau)| / n
\rightarrow 1/2$ as $n \rightarrow \infty$. That is, $\sigma \otimes
\tau$ is expected to share approximately half of its aligned pairs
with $\sigma$, and likewise with $\tau$. 
A proof of the above statement relies on the fact that the
expected number of cycles in a permutation is $\Theta(\log(n))$.  We
leave out further discussion on this, as it would require more basics
of abstract algebra, which is beyond the scope of this paper; see
\cite{Knuth1997,Dummit2004} for details.

\subsection{MAGNA: genetic algorithm-based GNA
  framework}\label{sect:MAGNA}

A genetic algorithm mimics the evolutionary process, guided by the
``survival of the fittest'' principle \cite{Back1996}.  It begins with
an initial ``population'' of a given number of ``members''.  Members
of a population ``crossover'' with one another to produce new members.
The ``child'' resulting from a crossover should resemble both of its
``parents''.  Crossing over different pairs of members at a given
generation yields new members, which comprise the new ``generation''
of members. The probability of a member being given a chance to
crossover with another member is determined by its ``fitness,'' so
that fitter members are more likely to crossover.  To prevent the size
of the population to grow without bound, the size is kept constant
across all generations, with only the fittest members surviving from
one generation to the next. To ensure that the maximum fitness of the
population is nondecreasing, with each generation, a designated
``elite'' class of the fittest members is automatically passed to the
next generation. As the algorithm progresses, newer generations are
produced, with fitness (hopefully) increasing, until a stopping
criterion is reached. To specify a genetic algorithm, we need to
specify all of the above parameters.

In MAGNA, members of a population are alignments.  We use different
types of initial populations: 1) all random alignments, 2) random
alignments mixed with an IsoRank's alignment, 3) random alignments
mixed with a MI-GRAAL's alignment, and 4) random alignments mixed with
a GHOST's alignment. Since we focus on \emph{topological} network
alignments (Section \ref{sect:motivation}), we produce all alignments
by using only topological information in the existing methods' node
scoring function. For each type of initial population, we test
populations of different sizes: 200, 500, 1,000, 2,500, 5,000, 10,000,
and 15,000.  (It is because the population sizes are large that we
cannot form an initial population consisting only of alignments
produced by an existing method, due to large computational complexity
of the existing methods.  Instead, we use in the initial population an
existing alignment and fill the remaining part of the population with
random alignments.)  The mathematical machinery from Section
\ref{sect:crossover} gives us a suitable crossover function for
producing a child alignment that resembles both of its parent
alignments. Our fitness function is the measure of alignment quality
we choose to optimize; in our case, it is edge correctness (EC),
induced conserved structure (ICS), or symmetric substructure score
(S$^3$) (Section \ref{sect:S3}), but it can be \emph{any} measure.  In
every generation, we keep the best half of the population from the
previous generation, and we fill the remaining half of the population
with alignments produced by crossovers. We select pairs of alignments
to be crossed as follows. At a given generation of population size
$p$, we have ${p \choose 2}$ crossover possibilities. This is too
large a number to consider all of them.  Thus, to select crossover
pairs, we use \emph{roulette wheel selection}, which is a commonly
adopted selection strategy for genetic algorithms \cite{Back1996}.
Roulette wheel selection chooses members with probability in linear
proportion to the members' fitness.  We let MAGNA run for many
generations.  We vary the number of generations from 0 to 2,000 in
increments of 200.  The fittest alignment from the last generation is
reported as the \emph{final} alignment. We describe MAGNA in the
pseudocode (Supplementary Algorithm S1). We provide MAGNA's
implementation upon request.

Our implementation of the alignment crossover function takes $O(|V|)$
time.  MAGNA's bottleneck, though, tends to be the computation of
alignment quality $F$. If the measure of $F$ is EC, ICS, or S$^3$,
then for a given alignment, it takes $O(|E|\log(|E|))$ time to compute
$F$.  Finally, sorting each generation of size $p$ takes $O(p\log(p))$
time, though this is typically negligible compared to the computation
of $F$. If MAGNA is run for $N$ generations, this brings the overall
time complexity of MAGNA to $O(N(p|V| + p|E| \log(|E|) + p \log(p)))$.
Note that most of MAGNA is embarrassingly parallelizable, which can
lead to a very high degree of speedup.

\subsection{New alignment quality measure}\label{sect:S3}
 
To motivate our new measure of alignment quality, the \emph{symmetric
  substructure score} (S$^3$), we first present drawbacks of existing
\emph{edge correctness} (EC) and \emph{induced conserved structure}
(ICS) measures.

Let $G_1(V_1, E_1)$ and $G_2(V_2, E_2)$ be two networks, and let $f :
V_1 \rightarrow V_2$ be an alignment between them. If $X \subseteq
V_2$, let $G_2[X]$ be the induced subnetwork of $G_2$ with node set
$X$. Also, if $H$ is a subnetwork of $G_2$, let $E(H)$ be its edge
set. Let $f(E_1) = \{(f(u), f(v)) \in E_2 : (u, v) \in E_1\}$, and let
$f(V_1) = \{ f(v) \in V_2 : v \in V_1\}$.

\begin{figure}[H]
\centering
\includegraphics[scale=0.6]{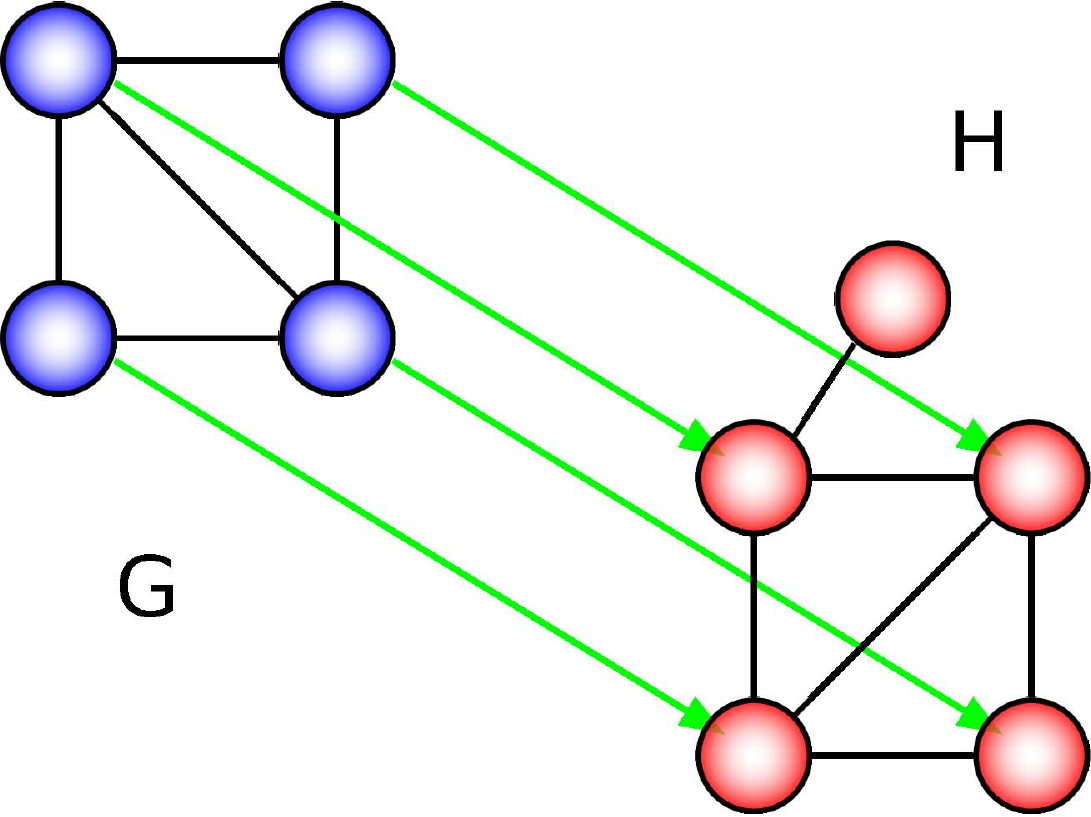}
\caption{Illustration of our new S$^3$ measure and its difference
    with EC and ICS. The illustrated alignment between nodes in
    network $G$ and nodes in network $H$ has an EC of $4/5=0.8$, an
    ICS of $4/5=0.8$, but an S$^3$ of $4/6=0.67$.  EC rewards for
    aligning four edges in $G$ to four edges in $H$ and penalizes for
    misaligning an edge in $G$ to a non-edge in $H$, but it fails to
    penalize for misaligning a non-edge in $G$ to an edge in $H$.
    Similarly, ICS rewards for aligning four edges in $G$ to four
    edges in $H$ and penalizes for misaligning an edge in $H$ (between
    the aligned nodes) to a non-edge in $G$, but it fails to penalize
    for misaligning an edge in $G$ to a non-edge in $H$.  Like EC and
    ICS, S$^3$ also rewards for aligning four edges in $G$ to four
    edges in $H$, but unlike EC or ICS, S$^3$ penalizes for
    misaligning \emph{both} an edge in $G$ to a non-edge in $H$ and a
    non-edge in $G$ to an edge in $H$.}\label{fig:S3}
\end{figure}

EC of $f$ is the ratio of the number of edges conserved by $f$ to the
number of edges in the source network: $\text{EC}(f) =
\dfrac{|f(E_1)|}{|E_1|}$ \cite{GRAAL}. Because EC is defined with
respect to the source but not the target network, it fails to penalize
alignments mapping sparser network regions to denser ones (Fig.
\ref{fig:S3}).

ICS of $f$ is the ratio of the number of edges conserved by $f$ to the
number of edges in the subnetwork of $G_2$ induced on the nodes in
$G_2$ that are aligned to the nodes in $G_1$: $\text{ICS}(f) =
\dfrac{|f(E_1)|}{|E(G_2[f(V_1)])|}$ \cite{GHOST}.  Because ICS is
defined with respect to the target but not the source network, it
fails to penalize alignments mapping denser network regions to sparser
ones (Fig.  \ref{fig:S3}).

Therefore, we define S$^3$ with respect to \emph{both} the source
network and the target network: $\text{S}^3(f) =
\dfrac{|f(E_1)|}{|E_1| + |E(G_2[f(V_1)])| - |f(E_1)|}$.  The
difference between EC, ICS, and S$^3$ is the denominator.
Intuitively, if $G_1$ and $G_2[f(V_1)]$ are overlaid into a composite
graph, then the denominator of S$^3$ is the number of unique edges in
this composite graph.  Thus, S$^3$ of an alignment is 100\% if and
only if $f$ is a \emph{perfect} embedding.  As such, S$^3$ penalizes
\emph{both} alignments that map denser network regions to sparser ones
and alignments that map sparser network regions to denser ones (Fig.
\ref{fig:S3}).

\section{Results and discussion}\label{sect:results_discussion}

\subsection{Validating MAGNA on  networks with \emph{known}
  node mapping}\label{sect:hc}

\subsubsection{Data description} 

We aim to validate MAGNA by analyzing the largest connected
component of the high-confidence yeast \emph{S.  cerevisiae} PPI
network \cite{Collins07} with 1,004 proteins and 8,323 PPIs.  We align
this network with the same network augmented with lower-confidence
PPIs from the same study \cite{Collins07}. We analyze different noise
levels, by adding 0\%, 5\%, 10\%, 15\%, 20\%, and 25\% of
lower-confidence PPIs; we add higher-scoring lower-confidence PPIs
first.  Since the networks being aligned are defined on the same set
of nodes and differ only in the number of edges, we know the correct
node mapping. An additional advantage of aligning these networks is
that the original is an exact subgraph of each noisy network.

\subsubsection{MAGNA parameters}

MAGNA requires several parameters: the type of initial population,
population size, maximum number of generations (i.e., iterations of
the genetic algorithm), and optimization function (i.e., alignment
quality measure) (Sections \ref{sect:MAGNA} and \ref{sect:S3}). We
evaluate MAGNA comprehensively and systematically, by varying values
of each parameter.

We use four different population types: random, IsoRank, MI-GRAAL, and
GHOST.  The random population aims to produce a high-quality alignment
from scratch (by relying only on our new alignment crossover
function), while the other three population types try to improve upon
the existing methods.  We test seven population sizes from 200 to
15,000.  We vary the maximum number of generations up to 2000, in
increments of 200.  We optimize three alignment quality measures: EC,
ICS, and S$^3$. See Sections \ref{sect:MAGNA} and \ref{sect:S3} for
details.

Each combination of initial population type, population size, maximum
number of generations, and optimization function results in one final
(best) alignment. This comprehensive testing has resulted in the total
of 5,544 \emph{final} alignments.

\vspace{0.1cm}

\hspace{-0.35cm}\textbf{The effect of the initial population type.}
Since we aim to compare MAGNA against IsoRank, MI-GRAAL, and GHOST (and
also random alignments), we continue by considering all four initial
population types and we discuss their effect on the alignment quality
in more detail below.

\vspace{0.1cm}

\hspace{-0.35cm}\textbf{The effect of population size.} We find that,
in general, larger population size is always preferred, independent of
the initial population type, maximum number of generations, and
optimization measure (Supplementary Section S1 and Supplementary Fig.
S1). Henceforth, we continue with the largest population size of
15,000.

\vspace{0.1cm}

\hspace{-0.35cm}\textbf{The effect of the maximum number of
  generations.}  We find that, in general, the larger the population
size, the larger number of generations is preferred, which is
$\sim$2,000 for random initial population, independent on the
optimization measure, and $\sim$400-1,200 for IsoRank, MI-GRAAL, or
GHOST initial population, depending on the optimization measure
(Supplementary Section S1 and Supplementary Fig. S1). In general,
GHOST initial population ``converges'' faster than MI-GRAAL and
IsoRank populations. Because of the design of MAGNA, the alignment
quality never drops from one generation to the following one. Thus,
even with IsoRank, MI-GRAAL, and GHOST populations, the results are
never worse at the 2,000$^{th}$ generation compared to the
400-1,200$^{th}$ generation. Thus, henceforth, we continue with the
maximum number of generations of 2,000, since this helps for at least
one population type without harming others. However, it is encouraging
that some methods can converge very fast, indicating that MAGNA can
produce high-quality alignments in short time.

\vspace{0.1cm}

\hspace{-0.35cm}\textbf{The effect of the optimization measure.} Since
we aim to compare our new S$^3$ measure with existing EC and ICS
measures, we continue by considering all three and we discuss their
effect on the alignment quality in more detail below.

\subsubsection{MAGNA evaluation and comparison with existing methods}

For each of the six noise levels, four initial population types (each
of size 15,000), and three optimization measures, we obtain with MAGNA
one final alignment, i.e., the best alignment (with respect to the
given optimization measure) at the 2,000$^{th}$ generation. In
addition, we study the original alignments produced by the existing
methods. Then, we compare these original alignments to those produced
by MAGNA to see whether MAGNA improves the alignment quality of the
existing alignments. Note that independent on which of the three
alignment quality measures (EC, ICS, or S$^3$) we optimize, the
question remains on how to best evaluate the correctness of the
resulting final alignment. Certainly, we could use \emph{any} of the
three alignment quality measures for this purpose. However, since the
true node mapping is known when aligning the high-confidence yeast PPI
network to its noisy counterparts, we can actually evaluate each
method more fairly by counting the number of correctly aligned node
pairs (or ``node correctness'').

\begin{figure}[H]
\centering
\includegraphics[scale=0.6]{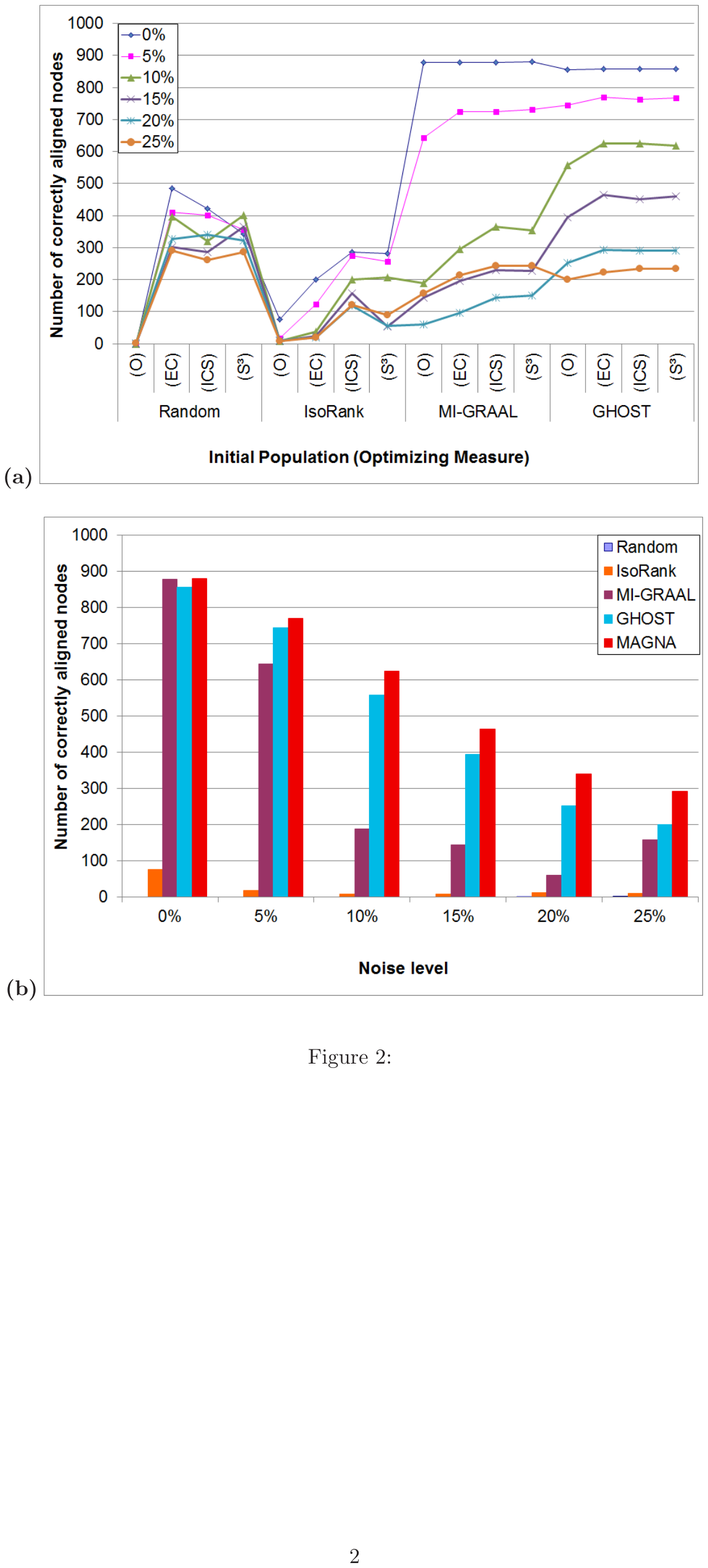}
\caption{ Correctness of alignments produced on noisy yeast networks
    (for noise levels in 0\%-25\% range), with respect to the number
    of correctly aligned node pairs.  Panel \textbf{(a)} shows results
    for alignments produced by four existing algorithms (Random,
    IsoRank, MI-GRAAL, and GHOST) as well as by running MAGNA on
    populations containing the alignments produced by the existing
    algorithms. We use four different populations, corresponding to
    the four existing algorithms. For each population, we show results
    for an original alignment produced by the existing algorithm (O),
    as well as for MAGNA's alignments produced when optimizing each of
    the following: EC, ICS, and $S^3$.  All results are for population
    size of 15,000 and for 2,000 generations.  Correctness of the
    alignments with respect to additional criteria, including EC, ICS,
    and S$^3$, are shown in Supplementary Fig. S2 and S3. Panel
    \textbf{(b)} shows, for each noise level, comparison of results
    from panel (a) between MAGNA's best alignment (over all initial
    population types and optimization measures) and the original
    alignments of the existing methods. (In most cases, original
    random alignments have scores close to 0 and are thus not
    visible.)  }\label{fig:hc}
\end{figure}

When we do this, we find that MAGNA improves \emph{all} of the original
alignments (i.e., all of the existing methods), across all levels of
noise, and for each of the three optimization measures (Fig.
\ref{fig:hc}). If we compute the ``improvement'' of MAGNA over an
existing method as the ratio of MAGNA's node correctness to the
existing method's node correctness, then MAGNA's improvement is up to
2,588\% upon IsoRank, up to 256\% upon MI-GRAAL, and up to 118\% upon
GHOST, depending on the noise level and optimization measure.  In
general, the higher the noise level, the larger our improvements upon
the existing methods (Fig.
\ref{fig:hc}).

\vspace{0.1cm}

\hspace{-0.35cm}\textbf{The effect of the initial population type.}
GHOST's \emph{original} alignments are overall slightly superior or
comparable to MI-GRAAL's original alignments, depending on the noise
level and the optimization measure, both are superior to IsoRank's
original alignments, and all three are superior to random original
alignments (Fig.  \ref{fig:hc}). These results are consistent to those
in the existing literature \cite{GRAAL,MIGRAAL,GHOST}. Thus, one might
expect that \emph{MAGNA's improved} alignments of GHOST would be of
better quality than MAGNA's improved alignments of MI-GRAAL, that both
would be of higher quality than MAGNA's improved alignments of IsoRank,
and that all three would be of higher quality than MAGNA's improved
alignments of random alignments. However, interestingly, we find that
this is not always the case (Fig.  \ref{fig:hc}).  Actually, in many
cases, there are surprising effects of the choice of the initial
population type.  For example, our improved alignments of MI-GRAAL are
sometimes better than our improved alignments of GHOST.  Or, even more
interestingly, for larger noise levels, it is the \emph{random}
population that results in the best alignments; that is, we improve
more when starting from completely random alignments than we do when
starting from the original alignments of IsoRank, MI-GRAAL, or GHOST
(Fig.  \ref{fig:hc}). More precisely, when we measure, for each of the
six noise levels, which initial population type results in the final
alignment with the highest node correctness score over all four
population types, we find that GHOST's initial population is the best
for three out of six noise levels (5\%, 10\%, and 15\%), random
initial population is the best for two noise levels (20\% and 25\%),
and MI-GRAAL's initial population is the best for the remaining noise
level (0\%).

The above results suggest that MAGNA is not only capable of improving
alignments generated by the existing methods, but it is also capable
of generating from completely random alignments its own new alignments
that are superior, especially for the higher noise levels. Interesting
implications are as follows.  First, because current PPI networks are
likely even noisier than those used in this section
\cite{Mering02,Venkatesan2009}, our results suggest that one might be
able to improve upon the current best PPI network alignments (of
different species, when the actual node mapping is unknown) simply
by using MAGNA on completely random alignments
of the PPI networks.  Second, recall that random initial population
converges the slowest of all populations, if at all.  And recall that
we stop MAGNA after 2,000 iterations, as all initial population types
but random one converge even before that.  Because of this, and
because random population is superior for larger noise levels, it is
possible that for such noise levels, the alignment quality could be
improved even further by running MAGNA longer, as dictated by the
available computing resources.

\vspace{0.1cm}

\hspace{-0.35cm}\textbf{The effect of the optimization measure.}  No
single optimization measure (out of EC, ICS, and S$^3$) is always
superior with respect to the node correctness as the alignment quality
measure; the results depend on the choice of MAGNA's parameters. Over
all noise levels, random initial population prefers (in the sense that
it results in the highest node correctness for) EC and S$^3$ equally,
IsoRank initial population prefers ICS, MI-GRAAL's initial population
prefers S$^3$, and GHOST initial population prefers EC (Fig.
\ref{fig:hc}).  Hence, S$^3$, as well as EC, seem to be preferred
overall in this context. Over all population types, four of the six
noise levels (5\%, 10\%, 15\%, and 25\%) prefer EC, one noise level
(0\%) prefers S$^3$, and the remaining noise level (20\%) prefers ICS.
Hence, EC seems to be preferred overall in this context.  We even
further study the effect of the three optimization measures by
computing Pearson correlation between the ``node correctness'' on one
hand and EC, ICS, or S$^3$ on the other, across all alignments from
Fig.  \ref{fig:hc}. A higher and more statistically significant
correlation would indicate that the given optimization measure is
capable of uncovering more correct alignments.  The node correctness
correlates the best and the most significantly with our new S$^3$
measure, suggesting its superiority over the existing measures (Table
\ref{tab:correlation}).

\vspace{0.2cm}

\begin{table}[htbp]
\begin{center}
\begin{tabular}{ccc}
\hline
measure & correlation & $p$-value\\
\hline
EC & 0.7538 & $3.9 \times 10^{-19}$ \\
ICS & 0.8339 & $2.7 \times 10^{-26}$ \\
S$^3$ & 0.8980 & $1.4 \times 10^{-35}$ \\
\hline
\end{tabular}
\end{center}
\vspace{-0.2cm}
\caption{Correlation between the node correctness and each of EC, ICS, and S$^3$.}\label{tab:correlation}
\end{table}

\section{Concluding remarks}

We present a conceptually novel framework for ``optimizing'' pairwise
global network alignment with respect to any alignment quality
measure, which outperforms the existing state-of-the-art methods.
Given the tremendous amounts of biological network data that are being
produced, network alignment will only continue to gain importance, as
it can be used to transfer biological knowledge from well
characterized species to poorly characterized ones between aligned
network regions.  Also, analogous to sequence alignment, network
alignment can be used to infer species' phylogeny based on
similarities of their biological networks.  Thus, it could lead to new
discoveries about the principles of life, evolution, disease, and
therapeutics.

\section*{Acknowledgements}

We thank Dr. H. Bunke for suggestions regarding the parameters of the
genetic algorithm and Drs. R. Patro and C. Kingsford for their
assistance with running GHOST. This work was funded by the NSF
CCF-1319469 and EAGER CCF-1243295 grants.

\end{document}